\documentclass[a4paper,12pt]{article}
\usepackage{graphicx}

\newcommand{\be}{\begin{equation}}
\newcommand{\ee}{\end{equation}}
\newcommand{\bea}{\begin{eqnarray}}
\newcommand{\eea}{\end{eqnarray}}
\newcommand{\beqn}{\begin{eqnarray}}
\newcommand{\eeqn}{\end{eqnarray}}
\newcommand{\ba}{\begin{array}}
\newcommand{\ea}{\end{array}}

\newcommand{\noi}{\noindent}
\newcommand{\grcl}{{\tt GRACE-loop$\;$}}
\newcommand{\ra}{\rightarrow}
\newcommand{\epem}{e^+ e^-}

\newcommand{\eettht}{$e^+ e^-\ra t \bar{t} H\;$}

\newcommand{\eennht}{$\epem \ra \nu \bar{\nu} H \;$}
\newcommand{\eennh}{$\epem \ra \nu \bar{\nu} H$}

\newcommand{\eezht}{$\epem \ra Z H \;$}

\newcommand{\eezhht}{$\epem \ra Z H H\;$}
\newcommand{\eeeeht}{$\epem \ra e^+ e^- H \;$}

\newcommand{\eeht}{$\epem \ra \epem H\;$}

\def\sm{${\cal{S}} {\cal{M}}\;$}
\def\sms{${\cal{S}} {\cal{M}}$}

\newcommand{\mz}{M_Z}
\newcommand{\mzz}{M_Z^2}

\newcommand{\ordalft}{${\cal O}(\alpha)\;$}

\newcommand{\beq}{\begin{equation}}
\newcommand{\eeq}{\end{equation}}

\newcommand{\sff}{s_{f\bar f}}
\newcommand{\calmz}{{\cal {M}}^{(0)}}
\newcommand{\calmo}{{\cal {M}}^{(1)}}
\newcommand{\calnz}{{\cal {N}}^{(0)}}
\newcommand{\gz}{\Gamma_Z}

\newcommand{\propz}{\sff-\mzz+i\gz \mz}

\newcommand{\propng}{\sff-\mzz}
\newcommand{\sts}{\tilde{\Pi}_T^{ZZ}(\sff)}
\newcommand{\stz}{\tilde{\Pi}_T^{ZZ}(M_Z^2)}
\newcommand{\azg}{A_{Z\gamma}\frac{\tilde{\Pi}_T^{Z\gamma}(\sff)}{\sff}}
\begin{document}

\begin{titlepage}

\vspace*{0.1cm}\rightline{KEK-CP-153}
\vspace*{0.1cm}\rightline{LAPTH-1049}
\vspace*{0.1cm}\rightline{TMCP-04-3}

\vspace{1mm}
\begin{center}

{\Large{\bf Electroweak corrections to Higgs production through
$ZZ$ fusion at the linear collider}}

\vspace{.5cm}

F.~Boudjema${}^{1)}$, J.~Fujimoto${}^{2)}$, T.~Ishikawa${}^{2)}$,
\\ T.~Kaneko${}^{2)}$, K.~Kato${}^{3)}$, Y.~Kurihara${}^{2)}$, Y.~Shimizu${}^{2)}$
and Y.~Yasui${}^{4)}$
\\

\vspace{4mm}

{\it 1) LAPTH${\;}^\dagger$, B.P.110, Annecy-le-Vieux F-74941,
France.}
\\ {\it
2) KEK, Oho 1-1, Tsukuba, Ibaraki 305--0801, Japan.} \\
{\it 3) Kogakuin University, Nishi-Shinjuku 1-24, Shinjuku, Tokyo
163--8677, Japan.} \\
 {\it 4) Tokyo Management College, Ichikawa,
Chiba,272-0001,Japan.}
\\

\vspace{10mm} \abstract{ We present the full ${{\cal O}}(\alpha)$
electroweak radiative corrections to \eeht .  The computation is
performed with the help of {\tt GRACE-loop}. The extraction of the
full QED corrections is performed, these are quite large at
threshold. The genuine weak corrections, for the linear collider
energies, when expressed in the $G_\mu$ scheme are of order $-2$
to $-4$\% for Higgs masses preferred by the latest precision data.
We also extract the $m_t^2$ type corrections and make a comparison
with the weak corrections for the process \eennht.}

\end{center}

\vspace*{\fill} $^\dagger${\small URA 14-36 du CNRS,
associ\'ee \`a l'Universit\'e de Savoie.} \normalsize
\end{titlepage}

\section{Introduction}
The hunt for the Higgs and the elucidation of the mechanism of
symmetry breaking is the primary task of all future colliders.
While the discovery of the Standard Model Higgs at the LHC has
been established for a wide range of Higgs masses, only rough
estimates on its properties will be possible, through measurements
on the couplings of the Higgs to fermions and gauge
bosons\cite{higgsproperties-lc-lhc} for example. Precise
extraction of these parameters will have to await the advent of
the Linear Collider, LC,
\cite{NLC-report,tesla-report,GLC-report}. Moreover for these
measurements to be confronted with theory, calculations beyond
tree-level are mandatory. Nonetheless, although the branching
ratios of the Higgs have, for some time now, been computed with
great precision it is only during the last year that the one-loop
electroweak radiative corrections to the main production channel
at the LC, \eennh, has been achieved\cite{eennhradcor2002,
eennhletter,Dennereennh1}. This is because this process is a
challenging $3$-body final state and was the first example of its
class to have received a full one-loop treatment.

\begin{figure*}[htbp]
\begin{center}
\includegraphics[width=16cm,height=9.5cm]{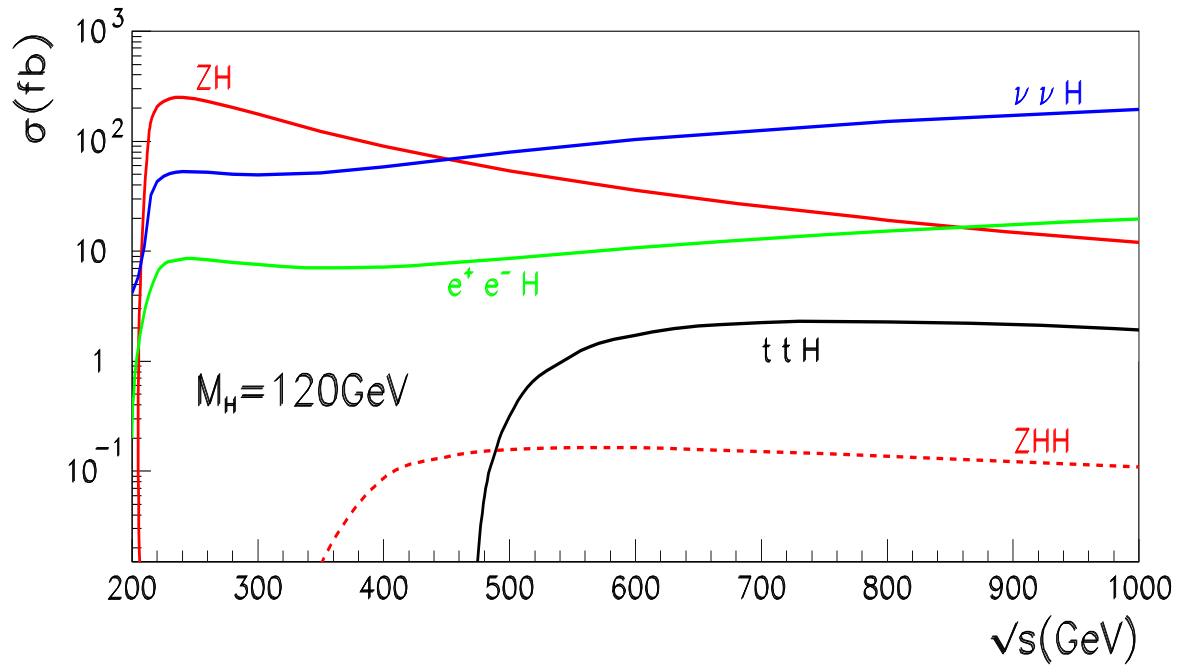}
\caption{{\em Total cross section for the main Higgs production
channels at the linear collider.}} \label{hprofile.fig}
\end{center}
\end{figure*}

As Fig.~\ref{hprofile.fig} shows, at the LC energies, \eennht by
far exceeds the yield of the oft-discussed two body process
\eezht. This is especially the case for Higgs masses in accord
with the latest indirect precision data\cite{mhlimit-03-2003} that
point towards a rather light Higgs with a mass not much greater
than the $W$ pair threshold. This is also within the range
predicted for the lightest Higgs of the minimal supersymmetric
model(MSSM). In this mass range at the LC, Fig.~\ref{hprofile.fig}
also makes evident that the process \eeht is a welcome addition
with a cross section in excess of the yield given by $ZH$
production around $1$TeV. It is also important to point out, that
especially at TeV energies, because the final electrons tend to be
lost in the beam pipe this reaction will have the same signature
as the single Higgs production in the $WW$ process so that in
effect one needs a precise determination of single Higgs
production. On the other hand, forcing the electrons to be
observable, this reaction can be used in improving the
determination of the $ZZH$ coupling\cite{ZZH-measure}. The aim of
this letter is to report on a full \ordalft calculation of this
process which has been missing so far thus completing a precise
knowledge of the full Higgs production profile at the LC. The
other production mechanisms shown in Fig.~\ref{hprofile.fig},
$e^+e^- \rightarrow t\bar{t}H$ which is crucial for the extraction
of the important $t\bar t H$ vertex, has been recently computed at
one-loop level \cite{eetthgrace,eetthdenner,eetthchinese}.
Even more recent, is the full \ordalft calculation of
\eezhht\cite{eezhhgrace,eezhhchinese} which is important for the
reconstruction of the Higgs potential through the measurement of
the $HHH$ vertex.

\section{{\tt Grace-loop} and the calculation of \eeht}
\subsection{Checks on the one-loop result}
The calculation of the complete electroweak corrections to the
process \eeht is performed with the help of {\tt GRACE-loop} which
is described in detail in\cite{nlgfatpaper}. This is a code for
the automatic generation and calculation of the full one-loop
electroweak radiative corrections in the \sms. The code has
successfully reproduced the results of a host of one-loop $2\ra 2$
electroweak processes\cite{nlgfatpaper}. During the past year {\tt
GRACE-loop}  provided the first results on the full one-loop
radiative corrections to $e^+ e^- \ra \nu \bar{\nu} H$
\cite{eennhradcor2002,eennhletter} and  on
\eettht\cite{eetthgrace} which have now been confirmed by an
independent group \cite{Dennereennh1,eetthdenner}. The calculation
we performed for the process \eezhht\cite{eezhhgrace} is also in
agreement with the calculation of\cite{eezhhchinese}. In \grcl we
adopt the on-shell renormalisation scheme according
to\cite{eennhletter,nlgfatpaper,kyotorc}.

For each calculation some stringent consistency checks are
performed. The results are verified by performing three kinds of
tests at some random points in phase space, that is before full
integration on phase space. For these tests to be passed one works
in quadruple precision. Details of how these tests are performed
are given in\cite{eennhletter,nlgfatpaper}.
Here we only recall the main features of these tests.\\

\noi {\it {\bf i)}}  We first check the ultraviolet finiteness of
the results. This test applies to the whole set of the virtual
one-loop diagrams. In order to conduct this test we regularise any
infrared divergence by giving the photon a fictitious mass (for
this calculation we set this at $\lambda=10^{-21}$GeV). In the
intermediate step of the symbolic calculation dealing with loop
integrals (in $n$-dimension), we extract the regulator constant
$C_{UV}=1/\varepsilon -\gamma_E+\log 4\pi$, $n=4-2 \varepsilon$,
and treat this as a parameter. The ultraviolet finiteness test is
performed by varying the dimensional regularisation parameter
$C_{UV}$. This parameter could then be set to $0$ in further
computation. Quantitatively for the process at hand, the
ultraviolet finiteness test gives a result that is stable over
$15$ digits when one varies the dimensional regularisation
parameter $C_{UV}$.\\

\noi {\it {\bf ii)}} The test on the infrared finiteness is
performed by including both the loop and the soft bremsstrahlung
contributions and checking that there is no dependence on the
fictitious photon mass $\lambda$. The soft bremsstrahlung part
consists of a soft photon contribution where the external photon
is required to have an energy $k_{\gamma}^0 < k_c\ll E_b$. $E_b$
is the beam energy. This part factorises and can be dealt with
analytically. For the QED infrared finiteness test we also find
results that are stable over $15$ digits when varying the
fictitious photon mass $\lambda$. \\

\noi {\it {\bf iii)}} A crucial test concerns the gauge parameter
independence of the results. Gauge parameter independence of the
result is performed through a set of five gauge fixing parameters.
For the latter a generalised non-linear gauge fixing
condition\cite{nlgfatpaper} has been chosen,
\beqn
\label{fullnonlineargauge} {{\cal L}}_{GF}&=&-\frac{1}{\xi_W}
|(\partial_\mu\;-\;i e \tilde{\alpha} A_\mu\;-\;ig c_W
\tilde{\beta} Z_\mu) W^{\mu +} + \xi_W \frac{g}{2}(v
+\tilde{\delta} H +i \tilde{\kappa} \chi_3)\chi^{+}|^{2} \nonumber \\
& &\;-\frac{1}{2 \xi_Z} (\partial.Z + \xi_Z \frac{g}{ 2 c_W}
(v+\tilde\varepsilon H) \chi_3)^2 \;-\frac{1}{2 \xi_A} (\partial.A
)^2 \;.
\eeqn
The $\chi$ represents the Goldstone. We take the 't Hooft-Feynman
gauge with $\xi_W=\xi_Z=\xi_A=1$ so that no ``longitudinal" term
in the gauge propagators contributes. Not only this makes the
expressions much simpler and avoids unnecessary large
cancellations, but it also avoids the need for higher tensor
structures in the loop integrals. The use of the five parameters,
$\tilde{\alpha}, \tilde{\beta}, \tilde{\delta}, \tilde{\kappa},
\tilde\varepsilon $ is not redundant as often these parameters
check complementary sets of diagrams.  Let us also point out that
when performing this check we keep the full set of diagrams
including couplings of the Goldstone and Higgs to the electron for
example, as will be done for the process under consideration. Only
at the stage of integrating over the phase space do we switch
these negligible contributions off. We should add that for this
test we omit all widths.  Here, the gauge parameter
independence checks give results that are stable over $21$ digits
(or better) when varying any of the non-linear gauge fixing
parameters. \\

\noi {\it {\bf iv)}} The tensor reduction, down to the scalar
integrals, of all loop integrals of rank $N<5$ is performed in the
space of the Feynman parameters as detailed in our
review\cite{nlgfatpaper}. For the $N=1,2$ scalar integrals we
implement full analytical formulae, whereas for the scalar $N=3,4$
integrals we use the {\tt FF} package\cite{ff} supplemented by our
own routines for the QED (infrared with photon-exchange) scalar
integrals. The treatment of the five-point functions is as
detailed  in our previous paper on
\eezhht\cite{eezhhgrace}.

\subsection{Input parameters}
The  input parameters for the calculation are the same as those we
have used for the calculation of   $e^+e^-\rightarrow Z H H$. We
recall them here.  We take for the fine structure constant in the
Thomson limit $\alpha^{-1}=137.0359895$ and  $M_Z=91.1876$GeV for
the $Z$ mass. The on-shell renormalisation program, which we have
described in detail elsewhere\cite{nlgfatpaper}, uses $M_W$ as an
input. However, the numerical value of $M_W$ is derived through
$\Delta r$\cite{Hiokideltar} with $G_\mu=1.16639\times 10^{-5}{\rm
GeV}^{-2}$\footnote{The routine we use to calculate $\Delta r$ is
the same as in our previous paper on \eezhht \cite{eezhhgrace}.}.
Thus, $M_W$ changes as a function of $M_H$. For the  lepton masses
we take $m_e=0.510999$ MeV, $m_\mu=105.658389$ MeV and
$m_\tau=1.7771$ GeV. For the quark masses, beside the top mass
$m_t=174$ GeV, we take the set $m_u=m_d=63$ MeV, $m_s=94$ MeV,
$m_c=1.5$ GeV and $m_b=4.7$ GeV. This set of effective quark
masses reproduces $\Delta \alpha^{(5)}_{\rm had}(M_Z^2)=276.094 \;
10^{-4}$ in a (naive) perturbative one-loop calculation of the
running $\alpha$ at $M_Z$, in excellent agreement with the value
used by the Electroweak Working Group\cite{mhlimit-03-2003}. With
this we find, for example, that $M_W=80.3766$GeV ($\Delta
r=2.549\%$) for $M_H=120$GeV and $M_W=80.3477$GeV ($\Delta
r=2.697\%$) for $M_H=180$GeV.
For this process, as we will discuss shortly we also need to
specify the $Z$-width. For this we have used a one-loop formula,
which gives $\Gamma_Z=2.4945$GeV for $M_H=120$GeV and
$\Gamma_Z=2.4927$GeV for $M_H=180$GeV. Both values are in fact
within $1\sigma$ of the experimental value $\Gamma_Z=2.4956$. The
lower bound of the \sm Higgs boson mass from LEP2 is 114.4
GeV\cite{mhlimit-direct}, while the indirect electroweak precision
measurement set an upper bound for the
\sm Higgs mass at about 200 GeV\cite{mhlimit-03-2003}. In this
paper, we therefore only consider a relatively light \sm Higgs
boson and take the illustrative values $M_H=120$GeV, $M_H=150$GeV
and $M_H=180$GeV.

\section{Tree-level calculation}
\begin{figure*}[hbtp]
\begin{center}
\includegraphics[width=10cm,height=3cm]{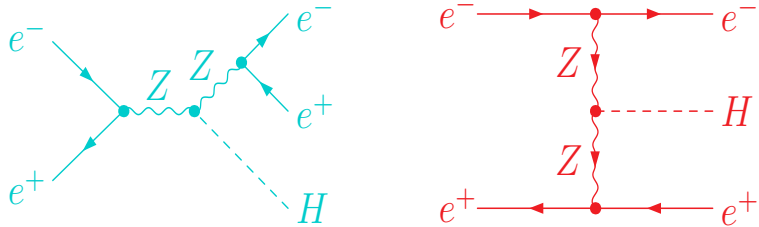}
\end{center}
\caption{{\em Contributing diagrams at tree-level in terms of the
s-channel type, left panel, obtained from $\epem \ra ZH$, and the
$t$-channel type from $ZZ$ fusion.}} \label{tree-level.fig}
\end{figure*}

\begin{figure*}[hbtp]
\begin{center}
\mbox{\includegraphics[width=0.5\textwidth,height=11cm]{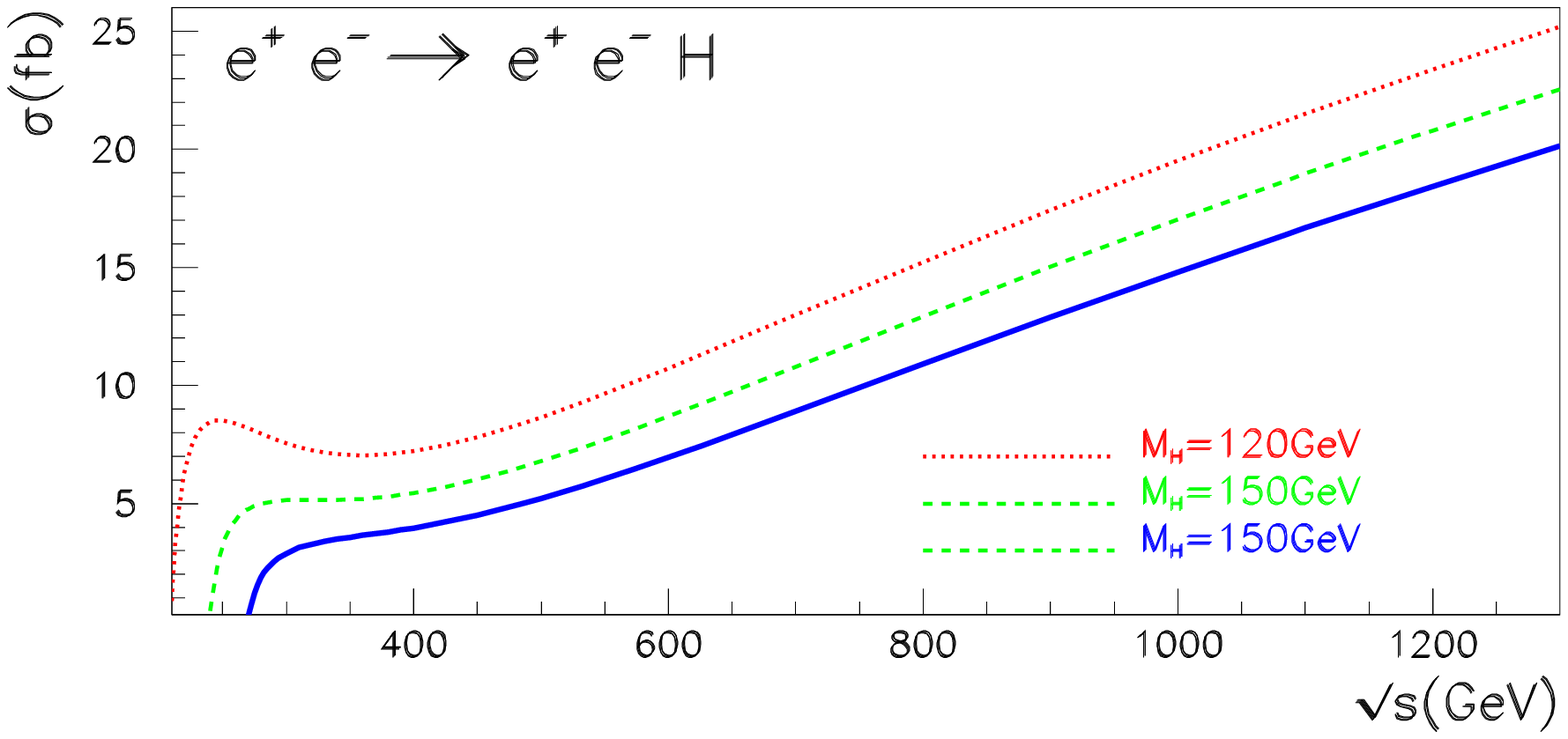}
\includegraphics[width=0.5\textwidth,height=11cm]{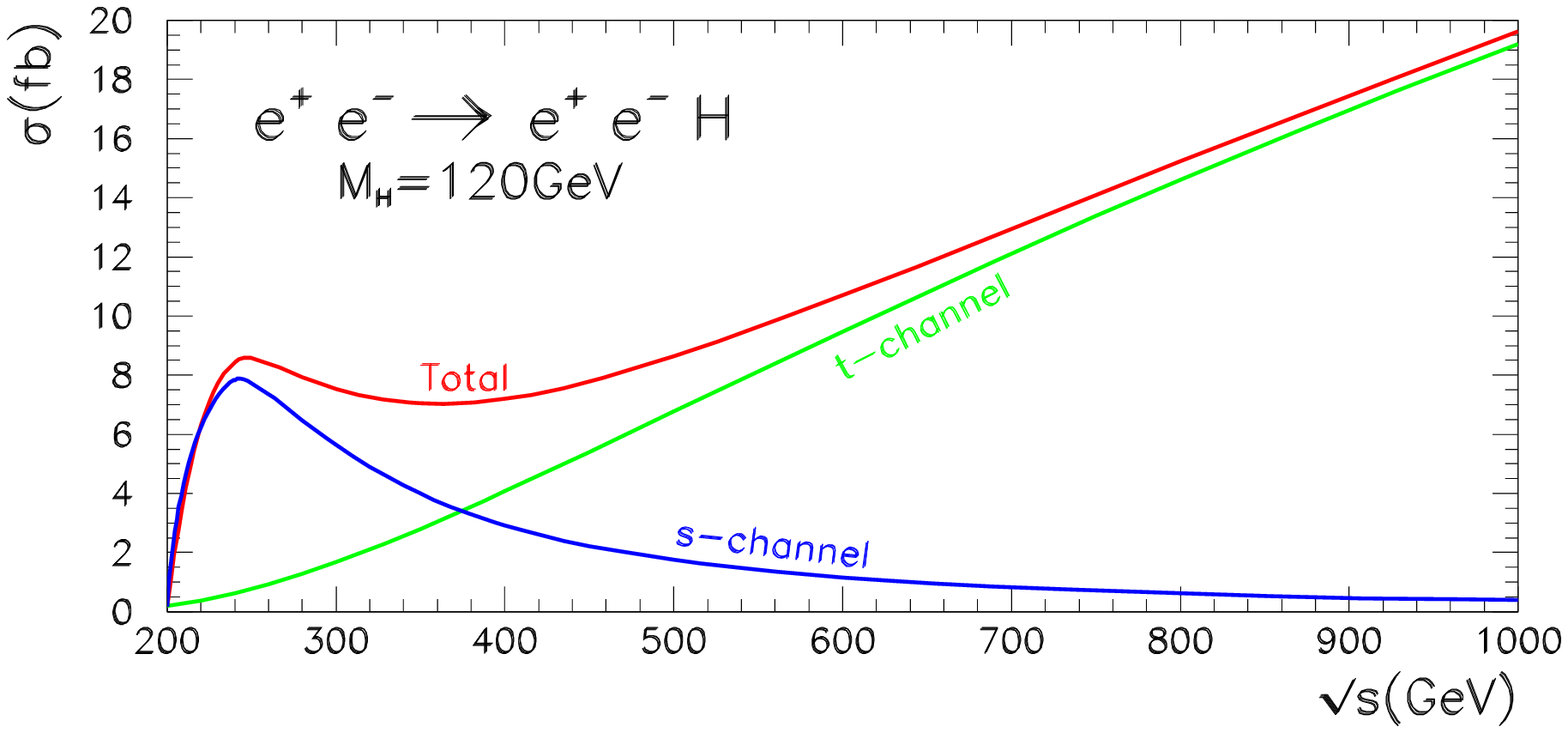}}

\vspace*{-3cm}
\mbox{\includegraphics[width=\textwidth,height=12cm]{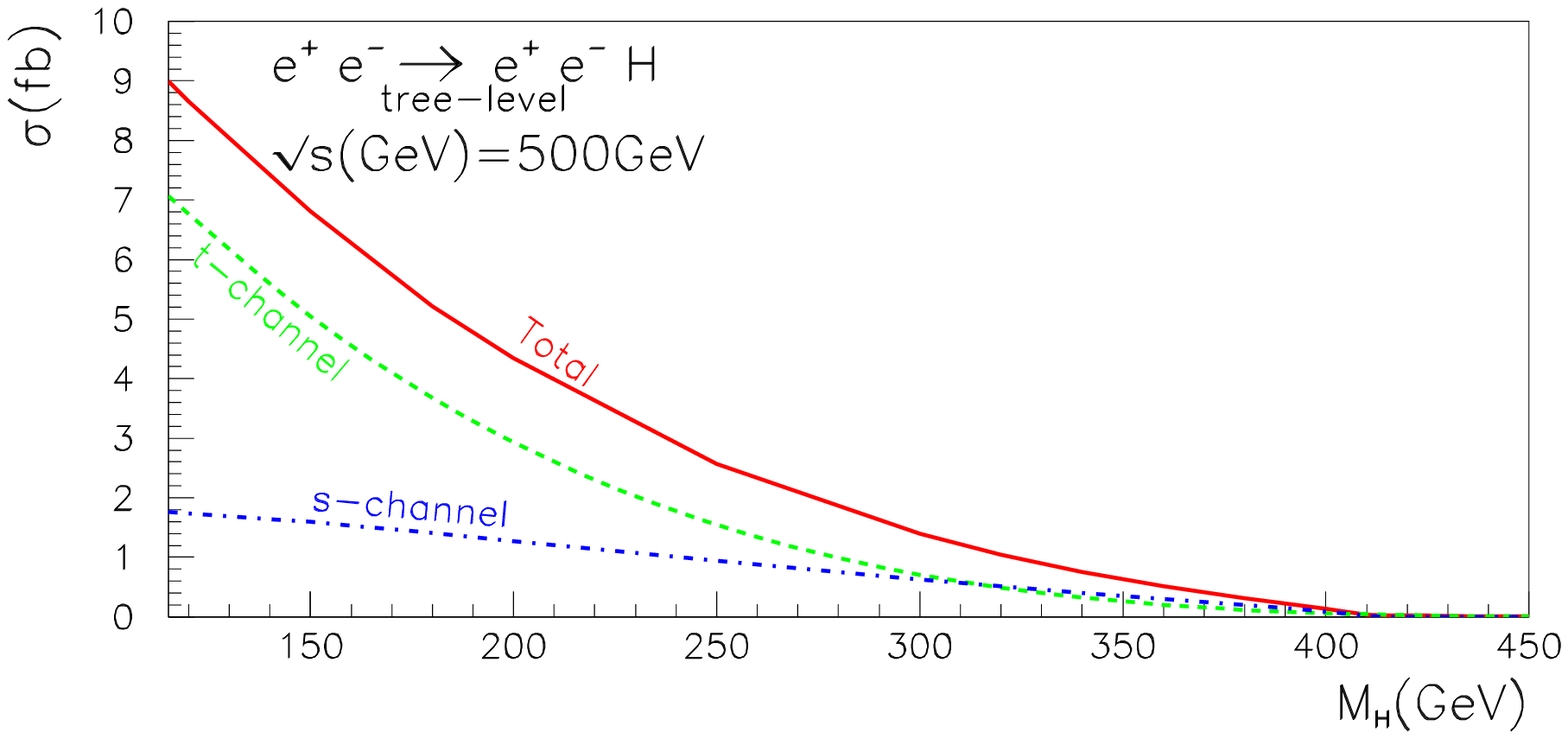}}
\raisebox{3.4cm}{ \mbox{ \hspace*{-10.cm}
\includegraphics[width=8.2cm,height=7.8cm]{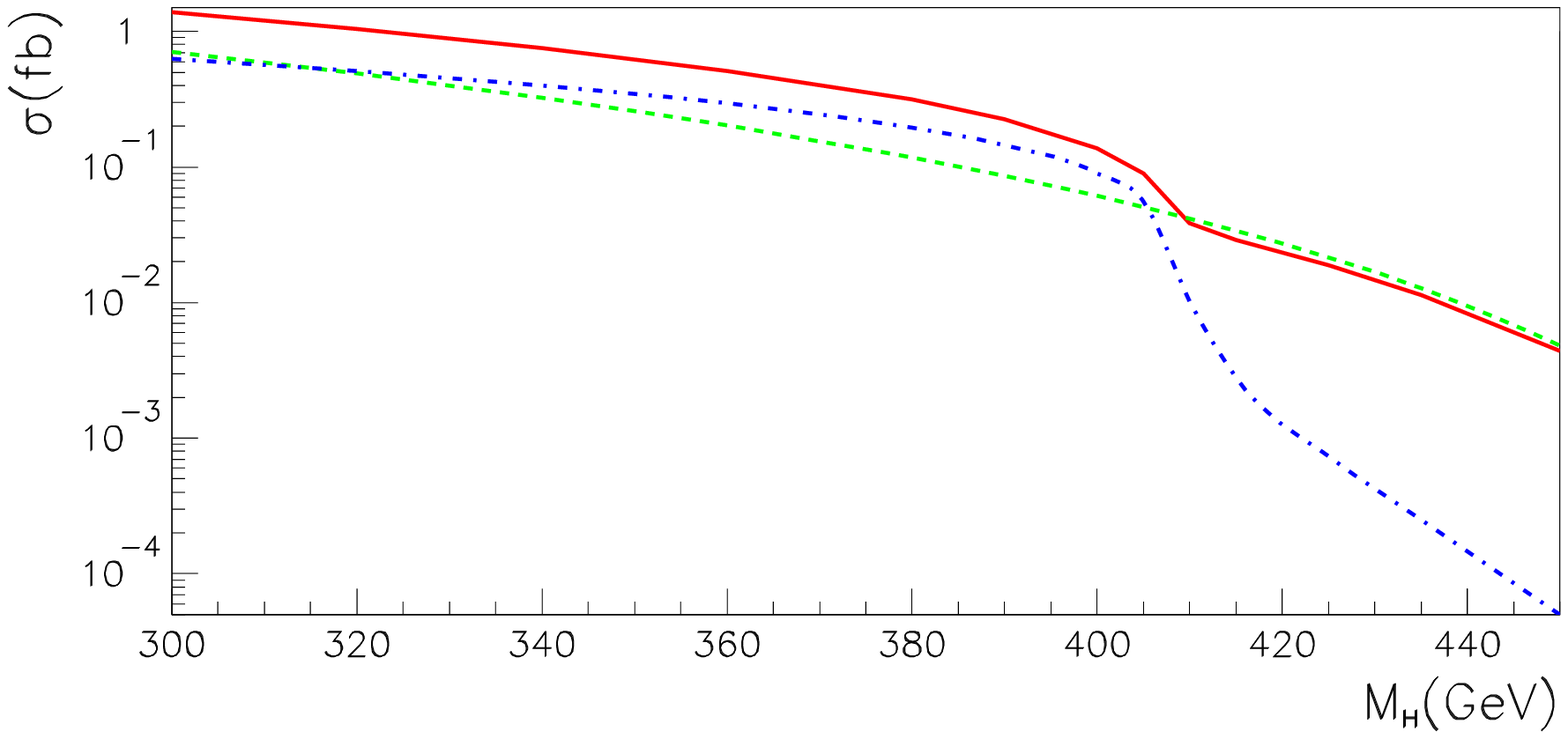}}}
\end{center}

\vspace*{-2cm} \caption{{\em Tree-level cross sections for \eeht.
The first panel shows the centre of mass energy dependence for
three value of the Higs mass. For $M_H=120$GeV, beside the total
cross section we show the $s$-channel and $t$-channel
contributions. In the panel at the bottom the dependence of the
cross section as a function of the Higgs mass is shown at a
centre-of-mass $\sqrt{s}=500$GeV.}} \label{tree-results}
\end{figure*}

At tree-level, in the unitary gauge, the \eeht process is built up
from $s$-channel diagram originating from \eezht and a $t$-channel
diagram which is a fusion type, see Fig.~\ref{tree-level.fig}.
Each type constitutes, on its own, a gauge independent process.
In fact the former (neglecting lepton masses) can be defined as
$\epem \ra \mu^+ \mu^- H$. In principle it is only the $Z$ coupling
to the outgoing lepton, in this $s$-channel contribution, which can
be resonating and thus requires a finite width. Nonetheless in our
code we dress both $Z$ in the $s$-channel type diagrams with a constant
$Z$ width. We apply no width to the $Z$ taking part in the
$ZZ$-fusion diagrams.\\
\noi For a light Higgs mass and for the LC energies, past
$\sqrt{s}=500$GeV, the cross section is dominated by the
$Z$-fusion, which grows (logarithmically) with energy, see
Fig.~\ref{tree-results}. The $s$-channel contribution follows
\eezht very closely, in the order of
$10\%$ of the total \eeht at $500$GeV but drops quite fast to
amount to a mere $2\%$ at $1$TeV for $M_H=120$GeV. With
$\sqrt{s}=500$GeV the cross section for $M_H=$120GeV is of the
order of $10$fb. Although it is about an order of magnitude
smaller than single Higgs production \eennht, this cross section
amounts to about $10^{4}$ events with a total integrated
luminosity of $1$ab$^{-1}$. The $1 \sigma$ statistical error
corresponds to about $1\%$. Thus the theoretical knowledge of the
cross section at $0.1\%$ is quite sufficient. At a moderate LC
energy of $\sqrt{s}=500$GeV, the cross section drops rather
quickly with increasing Higgs mass, Fig.~\ref{tree-results},
reaching the order of $1$fb for $M_H=300$GeV. This drop is
especially dramatic for the $t$-channel, for this Higgs mass at
this energy the $s$ and $t$-channel are about equal. The
$s$-channel slightly takes over before dropping precipitously at
threshold $\sqrt{s}\sim M_Z+M_H$.

It is also important to note, as this will help understand the QED
corrections, the steep increase of the cross section at threshold.

\section{Results at one-loop}
\subsection{Classification and overview of the diagrams}
\begin{figure*}[htbp]
\begin{center}
\includegraphics[width=16cm,height=14cm]{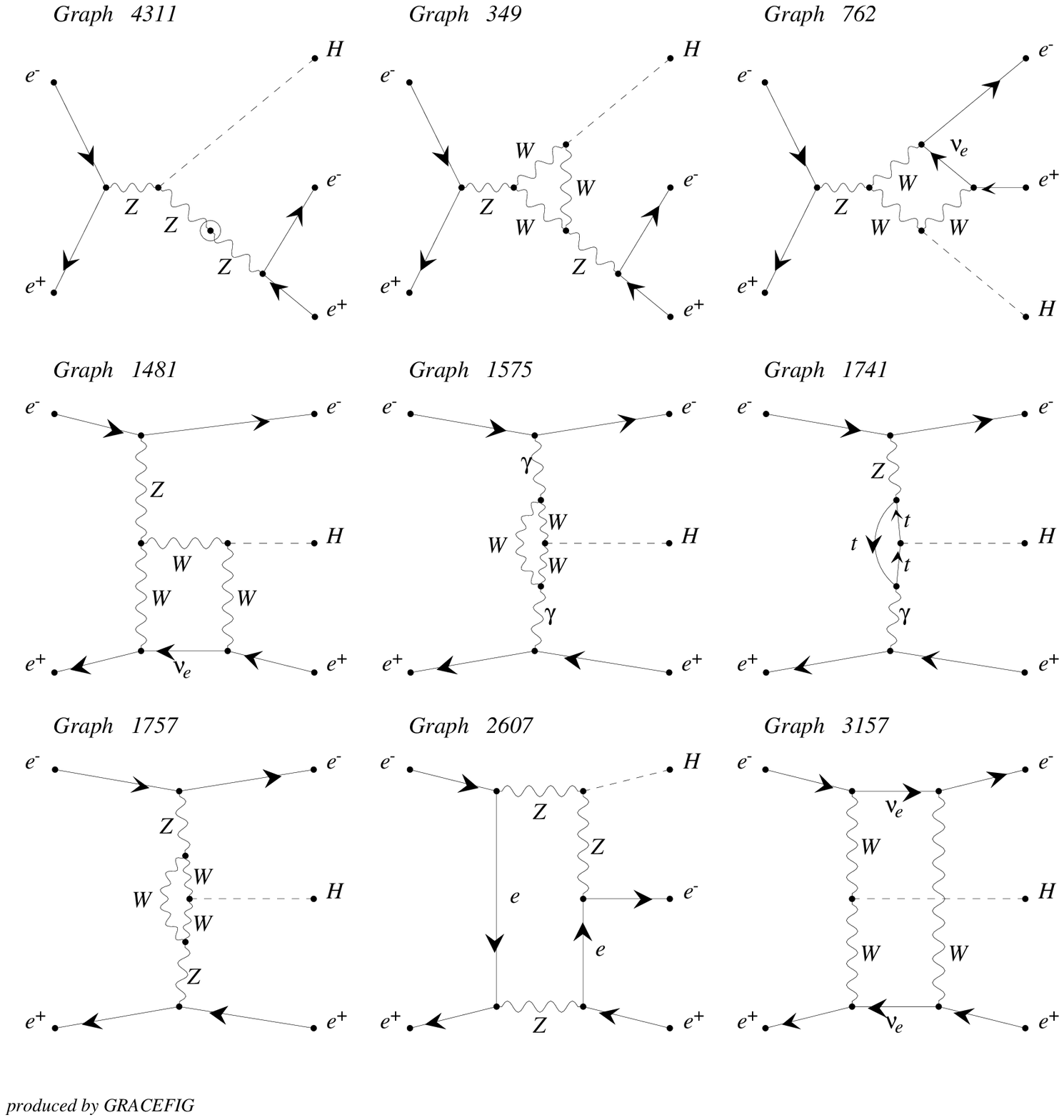}
\caption{\label{one-loop-diagrams} {\em A small selection of
different classes of loop diagrams contributing to \eeht. We keep
the same graph numbering as that produced by the system. {\tt
Graph 4311} belongs to the corrections from self-energies, here
both the virtual and counterterm contributions are generated and
counted as one diagram. {\tt Graph 349} shows a vertex correction.
Both graphs are considered as $s$-channel resonant Higgs-strahlung
contributions. {\tt Graph 762} represents a box correction, it is
a non resonant contribution, which  can not be deduced from
\eezht, but applies also to the correction to the $s$-channel
$\epem \ra \mu^+\mu^- H$. {\tt Graph 1481} is also a box
correction counted  as a correction to the $ZZ$ fusion. {\tt Graph
1575}, {\tt Graph 1741} and {\tt Graph 1757} are fusion type
corrections involving $\gamma \gamma$, $Z\gamma$ and $ZZ$ fusion.
{\tt Graph 2607} shows a pentagon correction which also counts as
an $s$-channel since it is induced for $\epem \ra \mu^+ \mu^- H$.
{\tt Graph 3157} on the other hand is a  pentagon correction that
only applies to \eeht \/.}}
\end{center}
\end{figure*}
The full set of the one-loop Feynman diagrams within the
non-linear gauge fixing condition consists of as many as $4470$
contributions for the full ${\cal O}(\alpha)$ correction. This is
to be compared to $1350$ diagrams for the \eennht process. A
selection of these diagrams is displayed in
Fig.~\ref{one-loop-diagrams}. These can be brought down to $510$
diagrams (as compared to $249$ for \eennht) when we neglect the
electron Yukawa coupling to the Higgs and the Goldstones. The
latter set we refer to as the production set as we use it for the
numerical computation of the cross section. The former set is used
to conduct the detailed tests of gauge parameter independence and
ultra-violet finiteness as described in the previous section. All
the one-loop diagrams can still be unambiguously divided into
$s$-channel and $t$-channel contribution. The former are the ones
one would still obtain in calculating $\epem \ra \mu^+ \mu^- H$
and obviously constitute a gauge invariant subset.\\
Of these one-loop diagrams, those corresponding to the pure QED
corrections can be easily extracted in a {\em gauge invariant}
way. They correspond to adding an internal photon between any two
external electron lines of the tree-level diagrams of
Figs~\ref{tree-level.fig}. These QED diagrams can therefore be
further subdivided into $s$- and $t$-channel contributions. These
QED corrections consist then of either pentagons or photonic
corrections to the $eeZ$ vertex (with its counterterms).

\subsection{Treatment of the $Z$-width at the loop-level}

\noi There is no definite, completely general and satisfactory
implementation of the width of an unstable gauge particle in loop
calculations. Many comparisons with different implementations of
the width\cite{fermionscheme,widthminami} have been made.
Nonetheless it has been found that applying a ``constant $Z$
width" is more appropriate than the running width and reproduces
the result of much more involved schemes (like the `fermion
scheme"\cite{fermionscheme}). For the case at hand, the
introduction of a width is required only for the $Z$ coupling to
the final electron pair. For this implementation of the width to
the $s$-channel neutral gauge boson, various ways should have a
negligible effect as has been found for the similar process
\eennht\cite{eennhletter,Dennereennh1}. Moreover the
$s$-channel contribution is much smaller than the $t$-channel
contributions, in the latter we do not endow the $Z$ propagators
with a width. Building up on the implementation of the width at
tree-level, we include a constant width to all $Z$ propagators
{\em not circulating in a loop} for the  $s$-channel type diagrams.
For example we add a width to all $Z$ in {\tt graphs 349,762,4311} of
Fig.~\ref{one-loop-diagrams}. As will be explained below when
discussing QED corrections, the $Z$ propagators inside QED pentagon
diagrams of the $s$-channel type are also calculated with the
addition of the constant $Z$-width.

For those one-loop diagrams with a self-energy correction to any
$Z$ propagator, represented by {\tt graph 4311} in
Fig.~\ref{one-loop-diagrams}, we follow a procedure along the
lines described in \cite{supplement100}. We will show how this is
done with a single $Z$ exchange coupling to a fermion pair of
invariant mass $\sff$.\\
\noi First, it is important to keep in mind, however, that our tree-level
calculation of the $s$-channel is done by supplying the width in
the $Z$ propagator. Therefore it somehow also includes parts of
the higher order corrections to the $Z$ self-energy which should
be subtracted when performing a higher order calculation. The case
at hand is as simple as consisting, at tree-level, of one single
diagram. The simplest way to exhibit this subtraction is to
rewrite, the zero-th order amplitude, $\calmz$, before inclusion
of a width, in terms of what we call the tree-level (regularised)
amplitude, ${\widetilde {\cal M}}^{{\rm tree}}$
\beqn
\label{calmz}
\calmz=\frac{\calnz}{\propng}=\overbrace{\frac{\calnz}
{\propz}}^{{\widetilde{\cal M}}^{{\rm tree}}}
\left(1+\frac{i(\gz^0+\Delta\gz)\mz}{\sff-\mzz}\right),
\quad \Delta\gz=\gz-\gz^0.
\eeqn
The $\gz^0$ contribution will be combined with the one-loop
correction while $\Delta\gz$ will be counted as being beyond
one-loop.

 At the one-loop level, before the summation {\em \`a la} Dyson and
the inclusion of any ``hard" width, the amplitude is gauge-invariant
and can be decomposed as
\beqn
\label{1loopgi}
\label{calmo} \calmo&=&\frac{\calnz}{\propng}
\frac{\sts}{\propng} +
\frac{\azg}{\propng}  + \frac{R_Z}{\propng} + C \nonumber \\
&=& \frac{1}{\propng}\left\{ \calnz \frac{\sts}{\propng} +
\left(\azg+R_Z\right) +(\sff-\mzz) C  \right\}.
\eeqn
The different contributions in $\calmo$ are the following. The
first term proportional to the tree-level contribution is due to
the renormalised transverse part of the $Z$ self-energy correction
$\tilde{\Pi}_T^{ZZ}$, including counterterms. Such a transition is
shown in {\tt Graph 4311} of Fig.~4. The term proportional to
$\azg$ comes from the renormalised transverse part of the
$Z$-$\gamma$ self-energy, $\tilde{\Pi}_T^{Z\gamma}$,  with the
photon attaching to the final fermion (this type is absent for
neutrinos in \eennh). The $R_Z$ terms combine one-loop corrections
but which nevertheless still exhibit a $Z$-exchange that couples
to the final fermions and hence these type of diagrams can be
resonant, an example is {\tt Graph 349} of Fig.~4. We can write
$R_Z=Z_{ZH}+V_Z^f$, where $Z_{ZH}$ corresponds to the part containing
the correction to $\epem \ra Z^\star H$, while $V_Z^f$ contains
the corrections to the final $Z_{f\bar f}$ vertex.
$Z_{ZH}(s_{f\bar f}=M_Z^2)$ corresponds to $\epem \ra Z H$ and is
gauge invariant at the pole. The term $C$ contains all the rest
which are apparently non-resonant\footnote{Strictly speaking we,
here, deal only with the pure weak corrections. In the infrared limit
some of the QED diagrams can be resonant and require a $Z$ width even
in a loop. This is discussed below.}, an example here is {\tt Graph
762} of Fig.~4. Both $\calmo$ and $\calmz$ are gauge invariant.

Our procedure, in effects, amounts to first regularising the
overall propagator in Eq.~\ref{1loopgi} by the implementation of a
constant $Z$ width and then combining the renormalised $Z$
self-energy part in Eq.~\ref{1loopgi} with the $\gz^0$ part of
Eq.~\ref{calmz}. Since our on-shell renormalisation procedure is
such that $Re\tilde{\Pi}_T^{ZZ}(M_Z^2)=0$ and since $\Gamma_Z^0=-Im
\Sigma_{T}^{ZZ}(M_Z^2)$, see \cite{nlgfatpaper}, our prescription
is to write
\beqn \label{width-rearrange}
\calmz+\calmo&\ra&\overbrace{\frac{\calnz}{\propz}}^{{\widetilde
{\cal M}}^{{\rm tree}}} + \overbrace{\frac{1}{\propz}
\widetilde{{\cal N}}^{1}}^{{\widetilde {\cal M}}^{{\rm 1-loop}}}
\nonumber \\
\widetilde{{\cal N}}^{1}&=&\calnz\frac{\left(\sts
-\stz\right)}{\propng} +
\left(\azg+R_Z\right) \nonumber \\
&+&(\sff-\mzz) C.\nonumber \\
\eeqn
The above prescription is nothing else but the factorisation
procedure avoiding double counting. It is gauge invariant but puts
the non-resonant terms to zero on resonance. In practice in the
automatic code, we supply a constant $Z$ width to all $Z$ not
circulating in a loop and by treating the one-loop $ZZ$
self-energy contribution as in Eq.~\ref{width-rearrange}. Up to
terms of order ${\cal{O}}(\Gamma_Z \alpha)$ this is equivalent to
Eq.~\ref{width-rearrange}. In particular the contribution of the
$C$ term  does not vanish on resonance, since its overall factor
is unity rather than the factor $(\sff-\mzz)/(\sff-\mzz+i\gz \mz)$
in the original factorisation prescription.

\subsection{Extraction of the QED corrections}
\begin{figure*}[bhtp]
\begin{center}
\includegraphics[width=0.8\textwidth,height=9cm]
{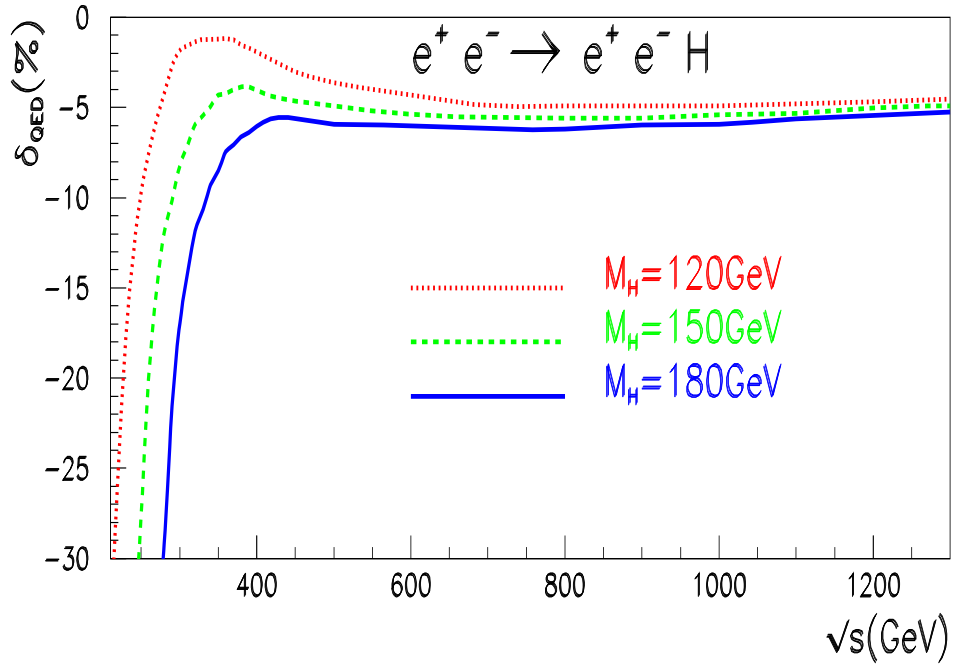}
\caption{{\em The purely QED corrections as function of the centre
of mass energy.}} \label{delta-qed.fig}
\end{center}
\end{figure*}
As already discussed, the QED corrections can be isolated
unambiguously in this process and separated into those belonging
to the $s$- and $t$-channel type corrections because of the
simple electric charge flow. For the $t$-channel contribution in
\eennht this was not the case thus preventing an unambiguous extraction
of the full QED correction. Only the leading log terms could be
identified there. Here, however, a new feature is that now one has
both initial state, final state and initial-final state
interference QED corrections. The last one in the $s$-channel consists
of pentagons. For this class of diagrams the infrared divergence occurs
also with the (final) $Z$ being resonant. We thus apply to this
internal $Z$ a constant width. This is akin to what occurs for
the $Z\gamma$ boxes in $\epem \ra f \bar f$ close to the $Z$
resonance\cite{supplement100}. In fact we recover this feature after
the decomposition of the pentagon, where part of this decomposition
maps precisely into the scalar $Z\gamma$ box. The $Z$ width is kept
consistently in the infrared $Z\gamma$ boxes. For such type of (scalar)
boxes we do not rely on the {\tt FF} package but revert to special
routines, see\cite{nlgfatpaper,supplement100}. After inclusion of the
bremsstrahlung contribution, this procedure shows that all
infrared divergences consistently cancel to a precision reaching
$15$ digits when we vary the fictitious photon mass.

The results of the purely QED corrections, including virtual, soft
and hard bremmstrahlung are displayed in Fig.~\ref{delta-qed.fig}
for the three Higgs masses $M_H=120,150,180$GeV. Around threshold
the corrections are large (and negative). For instance for
$\sqrt{s}=240$GeV and $M_H=120$GeV the total QED correction is
about $-13\%$ while for $M_H=180$GeV  it is about $-17\%$  at
$\sqrt{s}=300$GeV. It is to be noticed that the bulk of the
contribution for these energies and masses is due to the
$s$-channel process which exhibits a very sharp peak at threshold,
see Fig.~\ref{tree-results}. This explains the large negative
correction at small energies. This behaviour has been a feature of
all $s$-channel processes we have studied so far. As usual these
large initial state QED corrections could be resummed through, for
example, a structure function approach and QED parton shower, see
for instance\cite{qedps}. Past this energy range, to regions where
the $t$-channel contribution dominates and were the cross sections
are larger, the QED corrections vary slowly. In fact past about
$\sqrt{s}=500$GeV and up $1$TeV, the QED corrections almost
flatten out, at $1.3$TeV, they reach about $-5\%$ with little
Higgs mass dependence.

\subsection{Genuine electroweak corrections}
\begin{figure*}[hbtp]
\begin{center}
\includegraphics[width=16cm,height=10cm]{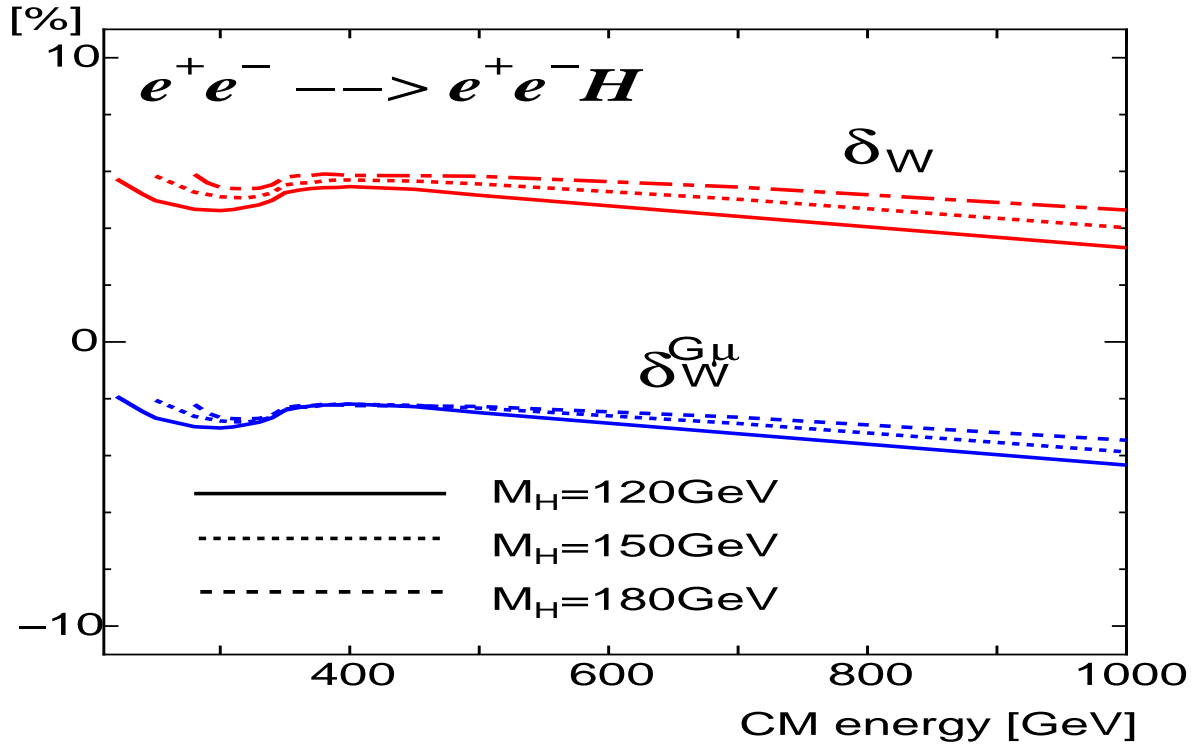}
\caption{{\em Genuine electroweak correction}} \label{dw.fig}
\end{center}
\end{figure*}
The most interesting part of the electroweak corrections consist
of the genuine weak corrections which in this process can be
unambiguously and easily extracted. Our results are shown in
Fig.~\ref{dw.fig} for the $3$ illustrative Higgs masses,
$M_H=120,150,180$GeV. In the $\alpha(0)$ scheme, the genuine weak
corrections for the entire process are positive. For small
energies just above threshold, one notices a feature we have
pointed out in our previous
calculations\cite{eennhletter,eetthgrace,eezhhgrace} and that we
termed as the {\em spoon-like} behaviour. This behaviour is
inherited from the \eezht process. The corrections show a very
slight dependence on the Higgs mass and very slowly decrease with
the center-of-mass energy, on the whole they are of order $\sim
5\%$. We have not attempted to extract the corrections for the
$s$-channel contribution as the latter follows the one we
conducted for the $s$-channel contribution to \eennht. Moreover at
the energies with largest cross sections, the most important
contribution is that of the $t$-channel. It is also possible to
extract part of the potentially large correction of order $m_t^2$,
as well as the logarithmic correction from small fermion masses in
the running of $\alpha$, by reverting to the $G_\mu$ scheme. In
the latter we simply trade $\alpha$ by $G_\mu$. This procedure
applied to this $2\ra 3$ process absorbs $3\Delta r$. The genuine
electroweak corrections for the entire process in the $G_\mu$
scheme are, in absolute terms, even smaller dropping from about
$-2\%$ at small energies to about $-4\%$ at $1$TeV, with again
very little dependence in the Higgs mass.

It is interesting to compare the genuine weak correction for this
process with the weak corrections we found for
\eennht\cite{eennhletter}. This is most appropriately done in the
$G_\mu$ scheme since in \cite{eennhletter} our input for the light
quarks was slightly different. One finds that the weak corrections
are almost of the same order in the two processes for the same
Higgs mass and the same energy. For instance for $M_H=150$GeV and
$\sqrt{s}=500$GeV, we find $\delta_W^G=-2.3\%$ in \eeeeht compared
to $\delta_W^G=-2.2\%$ in \eennht. At the same energy for
$M_H=180$GeV  we find $\delta_W^G=-2.3\%$ in \eeeeht compared to
$\delta_W^G=-1.9\%$ in \eennht. One should however exercise some
caution when interpreting this result which suggests that the
radiative corrections in the $G_\mu$ scheme in the two processes
are about equal. For example, corrections in both channels contain
other $m_t^2$ dependence that are not solely contained in $\Delta
r$. These  additional heavy $m_t^2$ corrections that could be
subtracted to arrive at an improved approximation are different
for the two processes. First of all as we had shown in
\cite{eennhletter} the genuine weak correction to the $s$-channel
contribution is rather large and can not be approximated by
subtracting these additional $m_t^2$ terms, in the heavy top mass
limit. On the other hand since at high energy the contribution of
the $s$-channel is negligible it is sensible to only parameterise
the corrections to the dominant $t$-channel. In \eennht these
heavy top mass $m_t^2$ terms originate from the $HWW$ vertex.
Subtraction of these  extra $m_t^2$ corrections defines, for the
dominant $t$-channel processes, an improved
approximation\cite{hwwapprox,Dennereennh1}

\beqn
\delta_{W,\nu \bar \nu H}^{imp} \sim \delta_W^G+5 X_t
\quad {\rm with } \quad X_t=\frac{\alpha}{16\pi} \frac{1}{s_W^2}
\frac{m_t^2}{M_W^2}.
\eeqn

For the $\epem H$ process the extra heavy mass $m_t^2$ dependence
comes from both the $ZHH$ vertex, with the same strength as in the
$WWH$ vertex, but also from the  $Z f
\bar{f}$ vertex as occurs in $Z\ra f \bar f$. Moreover this $m_t^2$
dependence is helicity dependent.

Let us recall that for the $ZZ$ fusion diagrams, the helicity
amplitudes that contribute are those of the type $e^{-}_\lambda
e^{+}_{\lambda^\prime} \ra e^{-}_\lambda e^{+}_{\lambda^\prime}$,
neglecting the electron mass. To make the notation simpler we will
write the helicity of the positron to correspond to the chirality
that represents its corresponding spinor. Pulling out the overall
left and right couplings $g_\lambda=g_{L,R}$, $ g_{L,R}= (-1+2
s_W^2, 2s_W^2)/(2 s_W c_W) $ where $c_W=M_W/M_Z$ and
$s_W^2=1-c_W^2$, we write the tree-level amplitudes as
\beqn
{\cal M}^{ZZ}_{\lambda,\lambda^\prime}=e^2 g_{\lambda}
g_{\lambda^\prime}\; {\cal N}^{ZZ}_{\lambda,\lambda^\prime}
\eeqn

The $m_t^2$ correction in the $G_\mu$ scheme to the helicity
amplitude writes as
\beqn
{\cal \delta M}^{ZZ}_{\lambda,\lambda^\prime}= {\cal
M}^{ZZ}_{\lambda,\lambda^\prime} \times
\delta^{G,t}_{W,\{\lambda,\lambda^\prime\}}; \quad
\delta^{G,t}_{W,\{\lambda,\lambda^\prime\}}=\frac{X_t}{2}\;\left(
1+\frac{6
c_W}{s_W}\left(g_\lambda^{-1}+g_{\lambda^\prime}^{-1}\right)\right)
\eeqn

At high energy the main contribution is from the $t$-channel where
both $Z$ are quasi on-shell. One expects, at high energy, that the
leading contribution is the same for all amplitudes ${\cal
N}^{ZZ}_{\lambda,\lambda^\prime}$. In this approximation, the
correction to the unpolarized cross section writes as
\beqn
\delta^{G,t}_{W}=X_t
\;\left(1+ 12 \frac{c_W}{s_W}
\frac{g_L+g_R}{g_L^2+g_R^2}\right)=X_t\;\left(1-48 c_W^2
\left(\frac{1-4s_W^2}{1+(1-4 s_W^2)^2}\right)\right)
\eeqn

The improved approximation for \eeeeht total cross section would
then write
\beqn
\delta_{W,\epem H}^{imp}&\sim&\delta_W^G-\delta^{G,t}_{W}
\eeqn

\begin{table}[h]
\begin{center}
\begin{tabular}{|c||c|c||c|c|}
\hline
 $\sqrt{s}=500GeV$ & \multicolumn{2}{c||}{\eennht} & \multicolumn{2}{c|}{\eeeeht} \\
\cline{2-5}
$M_H$(GeV)& $\delta_W^G(\%)$ & $\delta_W^{imp}(\%)$ & $\delta_W^G(\%)$ & $\delta_W^{imp}(\%)$ \\
120&  -2.3 & -0.8 & -2.5 & -1.6\\
 150 &   -2.2 & -0.7 & -2.3 & -1.4\\
  180 & -1.9 & -0.4 & -2.3 & -1.4\\ \hline
  \hline
$\sqrt{s}=1$TeV& \multicolumn{4}{|l|}{ }  \\
 \cline{2-5}
  $M_H=150$GeV & -3.6 & -2.1 & -3.9 & -3.0\\ \hline
  \hline
\end{tabular}
\end{center}
\caption{{\em Comparing the genuine weak corrections in the
$G_\mu$ scheme and after subtraction of $m_t^2$ terms between
\eennht and \eeeeht.}}\label{tab-eeh-nnh}
\end{table}

We see, Table~\ref{tab-eeh-nnh}, that the corrections for the
two-processes are within $1\%$ of each other in the improved
approximation, both for a centre of mass energy of $500$GeV and
$1TeV$ after subtracting the $m_t^2$ corrections. It is also
important to remark that   for the $\epem H$ process the full QED
corrections have been subtracted to define the genuine weak
corrections whereas for the neutrino process only the universal
QED corrections have been subtracted in \cite{eennhletter}. This
could be another reason for the fortuitous agreement between the
weak corrections expressed in the $G_\mu$ scheme for the two
processes. It is also interesting to note that especially for the
\eennht process the correction in the improved approximation for
the light Higgs masses we have studied is quite below $1\%$ at
$500$GeV. It is about $2\%$ at 1TeV. We note that the large
$m_t^2$ approximation works better when the top mass is much
heavier than all other scales in the problem. Though this is
really not the case compared to $M_W,M_Z$ and $M_H$, it is still a
relatively good approximation. We have however checked explicitly,
within the GRACE system, that the approximation works much better
with higher Higgs masses $m_t^2 \gg 174$GeV, while the accuracy of
the approximation is less clear in the energy range of the LC with
$m_t^2\ll s$. In any case, we should stress that considering the
accuracy with which the single Higgs cross sections will be
measured, our calculations of
\eeeeht and
\eennht show that one still needs to perform the full electroweak
corrections for a precision measurement of the cross section.

\section{Conclusions}
With the process \eeeeht we now have complete one-loop prediction
for the most dominant processes for Higgs production at the linear
collider. Considering that one of the primary aims of this machine
is a precise determination of the properties of the Higgs, such
calculations have for long been missing. For this process, we have
found that the genuine weak corrections when in expressed in the
$G_\mu$ scheme are quite moderate being of order $-2$ to $-4$\%
for the Higgs masses preferred by the present indirect limits. The
QED corrections in the energy region of LC is quite modest, though
it can be quite large at energies around the threshold of
production, where the cross sections are small anyhow.

\vspace{1cm}
\noi {\bf Acknowledgments} \\
\noi This work is part of a collaboration between the {\tt GRACE}
project in the Minami-Tateya group and LAPTH. We gratefully
acknowledge the participation of Genevi\`eve B\'elanger throughout
this project and  would like to thank her for her help and
comments. We would also like to thank Denis Perret-Gallix for his
continuous interest and encouragement. This work was supported in
part by the Japan Society for Promotion of Science under the
Grant-in-Aid for scientific Research B(N$^{{\rm o}}$ 14340081) and
PICS-GDRI 397 of the French National Centre for Scientific
Research (CNRS). We also thank {\tt IDRIS}, {\em Institut du
D\'evelopement et des Resources en Informatique Scientifique} for
the use of their computing resources (Project N$^{{\rm o}}$
041716).


\begin{thebibliography}{10}

\bibitem{higgsproperties-lc-lhc}
Precision Higgs Working Group of Snowmass 2001, J. Conway {\it et
al.},
  FERMILAB-CONF-01-442, SNOWMASS-2001-P1WG2, Mar 2002. 20pp. Contributed to
  APS/DPF/DPB Summer Study on the Future of Particle Physics (Snowmass 2001),
  Snowmass, Colorado, 30 Jun - 21 Jul 2001;hep-ph/0203206.

\bibitem{NLC-report}
T.~Abe {\it et al.} [American Linear Collider Working Group
Collaboration],
  ``Linear collider physics resource book for Snowmass 2001,'' in {\it Proc. of
  the APS/DPF/DPB Summer Study on the Future of Particle Physics (Snowmass
  2001) }; hep-ex/106055, hep-ex/106057, hep-ex/106058.

\bibitem{tesla-report}
J.~A.~Aguilar-Saavedra {\it et al.} [ECFA/DESY LC Physics Working
Group
  Collaboration], ``TESLA Technical Design Report Part III: Physics at an e+e-
  Linear Collider,'' arXiv:hep-ph/0106315.

\bibitem{GLC-report}
K.~Abe {\it et al.} [ACFA Linear Collider Working Group
Collaboration],
  ``Particle physics experiments at JLC,'' arXiv:hep-ph/0109166.

\bibitem{eennhradcor2002}
G. B\'{e}langer, F. Boudjema, J. Fujimoto, T. Ishikawa, T. Kaneko,
K. Kato and
  Y. Shimizu, Nucl.Phys. (Proc. Suppl.) {\bf 116} (2003) 353; hep-ph/0211268.

\bibitem{eennhletter}
G. B\'{e}langer, F. Boudjema, J. Fujimoto, T. Ishikawa, T. Kaneko,
K. Kato and
  Y. Shimizu, Phys. Lett. {\bf B559} (2003) 252; hep-ph/0212261.

\bibitem{Dennereennh1}
A.Denner, S.Dittmaier, M.Roth and M.M.Weber, Phys.Lett. {\bf B560}
(2003) 196;
  hep-ph/0301189 and Nucl.Phys. {\bf B660} (2003)289; hep-ph/0302198.

\bibitem{mhlimit-03-2003}
Martin W. Gr\"unewald, Invited talk presented at the Mini-Workshop
"Electroweak
  Precision Data and the Higgs Mass" DESY Zeuthen, Germany, February 28th to
  March 1st, 2003, hep-ex/0304023.

\bibitem{ZZH-measure}
For a recent study see, T.~Barklow,\\ {\tt
  http://www.slac.stanford.edu/$\sim$timb/talks/higgs$\_$1tev$\_$alcpg$\_$jan$%
\_$2004.pdf}.

\bibitem{eetthgrace}
G. B\'{e}langer, F. Boudjema, J. Fujimoto, T. Ishikawa, T. Kaneko,
K. Kato, Y.
  Shimizu and Y.~Yasui, Phys.Lett. {\bf B571} (2003) 163; hep-ph/0307029.

\bibitem{eetthdenner}
A.Denner, S.Dittmaier, M.Roth and M.M.Weber, Phys.Lett. {\bf B575}
(2003) 290;
  hep-ph/0307193 and hep-ph/0309274.

\bibitem{eetthchinese}
You Yu, Ma Wen-Gan, Chen Hui, Zhang Ren-You, Sun Yan-Bin and Hou
Hong-Sheng,
  Phys.Lett. {\bf B571} (2003) 85; hep-ph/0306036. 
  this paper for configurations around thresholds and 
  reproduce those of \cite{eetthgrace} and \cite{eetthdenner}, both of which
  agree 

\bibitem{eezhhgrace}
G. Belanger, F. Boudjema, J. Fujimoto, T. Ishikawa, T. Kaneko, Y.
Kurihara, K.
  Kato, and Y. Shimizu, Phys. Lett. {\bf B576} (2003) 152; hep-ph/0309010.

\bibitem{eezhhchinese}
Zhang Ren-You, Ma Wen-Gan, Chen Hui, Sun Yan-Bin, Hou Hong-Sheng,
Phys.Lett.
  {\bf B578} (2004) 349; hep-ph/0308203.

\bibitem{nlgfatpaper}
G. B\'{e}langer, F. Boudjema, J. Fujimoto, T. Ishikawa, T. Kaneko,
K. Kato and
  Y. Shimizu, hep-ph/0308080.

\bibitem{kyotorc}
K.~Aoki, Z.~Hioki, R.~Kawabe, M.~Konuma and T.~Muta, Suppl. Prog.
Theor. Phys.
  {\bf 73} (1982) 1.

\bibitem{ff}
G. J. van Oldenborgh , Comput. Phys. Commun. {\bf 58} (1991) 1.

\bibitem{Hiokideltar}
We use the code from Z. Hioki, see for example Z.Hioki, Zeit.
Phys. C49 (1991),
  287, see also Z. Hioki, Acta Phys.Polon. {\bf B27} (1996) 2573;
  hep-ph/9510269.

\bibitem{mhlimit-direct}
The LEP Higgs Working Group, \\
  http://lephiggs.web.cern.ch/LEPHIGGS/www/Welcome.html.

\bibitem{fermionscheme}
W.~Beenakker {\it et al.,} Nucl.Phys. {\bf B500} (1997) 255.

\bibitem{widthminami}
Y.~Kurihara, D.~Perret-Gallix and Y.~Shimizu Phys.Lett. {\bf B349}
(1995) 367.

\bibitem{supplement100}
J.~Fujimoto, M.~Igarashi, N.~Nakazawa, Y.~Shimizu and
K.~Tobimatsu, Suppl.
  Prog.~Theor.~Phys. {\bf 100} (1990) 1.

\bibitem{qedps}
T.~Munehisa, J.~Fujimoto, Y.~Kurihara and Y.~Shimizu, Prog. Theor.
Phys. {\bf
  95} (1996) 375; hep-ph/9603322.\\ Y~.Kurihara, J. Fujimoto, T. Munehisa and
  Y. Shimizu, Prog.Theor.Phys. {\bf 96} (1996) 1223.

\bibitem{hwwapprox}
For a review see, B. A. Kniehl, Int.J.Mod.Phys. {\bf A17} (2002)
1457.

\end{thebibliography}
\end{document}